\documentstyle[11pt,aaspp4]{article}
 
\begin{document}

\title{OH SATELLITE LINE MASERS IN THE NUCLEUS OF NGC~253}

\author{D. T. FRAYER and E. R. SEAQUIST }\affil{Astronomy
  Department, University of Toronto, Toronto, ON, M5S 3H8, Canada}
\author{AND} 
\author{D. A. FRAIL}\affil{National Radio Astronomy Observatory,
  P.O. Box 0, Socorro, NM  87801}

\begin{abstract}

  We report the detection of 1720 and 1612 MHz OH ground--state
  satellite line maser regions in the starburst galaxy NGC 253.  We find
  ten compact maser features all of which are located along the narrow
  ridge of bright radio continuum emission within the central 10 square
  arcseconds of the nucleus.  The compact OH features appear physically
  associated with regions of dense molecular gas, supernova remnants,
  and HII regions.  The data for several of the features are consistent
  with collisional inversion from the interaction of supernova remnant
  shocks with molecular gas, similar to that found for Galactic
  supernova remnants (Frail et al. 1996).  For many other features near
  the nucleus of NGC~253, we find conjugate behavior in the 1720 and
  1612 MHz lines, with one line showing stimulated emission and the
  other showing stimulated absorption.  These results suggest the
  importance of infrared pumping associated with the active star
  formation in NGC~253 and is consistent with large column densities of
  molecular gas (N[H$_2] \ga 10^{22}$~cm$^{-2}$).

\end{abstract}


\section{INTRODUCTION}

To date, the observations of extragalactic OH masers have been mainly
associated with the 1667 and 1665 MHz main--line OH transitions.  The
very first detections of extragalactic molecules were made in absorption
for these transitions in NGC~253 and M82 (Weliachew 1971).  Subsequent
observations of NGC~253 revealed a strong narrow 1667 MHz emission line
which was explained as maser amplification of the nuclear continuum
source (Whiteoak \& Gardner 1973).  Extremely luminous 1667 and 1665 MHz
OH megamasers ($L \sim 10^{3} L_{\odot}$) have since been found in
several bright IRAS galaxies (Baan, Wood, \& Haschick 1982; Mirabel \&
Sanders 1987; Kazes \& Baan 1991; Staveley--Smith et al. 1992) and are
believed to result from the pumping of the nuclear molecular gas by
far--infrared photons (Baan 1989; Randell et al. 1995).

Although the 1720 and 1612 MHz satellite--line transitions have received
considerably less attention than the main--line transitions, new
research indicates that the satellite--lines may provide an additional
diagnostic of the nuclear regions in starburst galaxies.  Recently,
Frail et al. (1996) have found a population of 1720 MHz OH masers
associated with Galactic supernova remnants (SNRs).  The passage of SNR
shocks through dense molecular clouds ($10^{3}{\rm cm}^{-3} < n({\rm
H}_2) < 10^{5}{\rm cm}^{-3}$) is expected to produce strong 1720 MHz
masers via collisional pumping (Elitzur 1976).  For starburst galaxies
which have large concentrations of molecular gas and numerous supernova
remnants, we could expect to find a significant population of 1720 MHz
masers.  Also for active galaxies, we could expect to detect conjugate
emission and absorption in the 1720 and 1612~MHz lines.  Such behavior
has recently been reported for Centaurus~A and indicates the pumping of
large column densities of molecular gas from a strong source of
far--infrared radiation (van Langevelde et al. 1995).  Sources showing
similar behavior in other galaxies could, perhaps, be a signpost for the
AGN with a molecular gas torus, as suggested for the 1612~MHz OH maser
source in M82 (Seaquist, Frayer, \& Frail 1997).

In this paper we report on VLA high resolution 1720 and 1612~MHz
satellite--line OH observations of the nearby starburst galaxy NGC~253.
NGC 253 is a nearby (D$\sim$ 3 Mpc) edge--on spiral galaxy which has an
unusually high rate of star formation in the central nuclear region
(Rieke et al. 1980).  Radio continuum observations have revealed a large
population of compact sources, presumably supernovae, SNRs, and HII
regions associated with the star formation (Turner \& Ho 1985; Antonucci
\& Ulvestad 1988; Ulvestad \& Antonucci 1991).  Recently, Ulvestad \&
Antonucci (1997) [UA97] have compiled a list of 64 distinct compact
sources using subarcsecond radio observations to resolve many of the
sources.  These radio observations suggest that about half of the
sources are dominated by thermal emission associated with HII regions,
and half are SNRs dominated by nonthermal emission (UA97).  The
distribution of nuclear SNRs and HII regions is similar to that of the
brightest HCN (Paglione, Tosaki, \& Jackson 1995) and CS (Peng et
al. 1996) emission regions in NGC~253, implying that the nuclear star
formation is associated with the dense molecular gas.  The typical
densities estimated from the HCN and CS observations are $n({\rm H}_2)
\sim 10^{4} - 10^{5}~{\rm cm}^{-3}$, and the total mass of the dense
molecular gas exceeds $10^{8} M_{\sun}$ and accounts for 17\% of the
dynamical mass in the central regions of NGC~253 (Peng et al. 1996).
The combination of a large number of SNRs and a massive component of
dense molecular gas suggests that NGC~253 is an ideal candidate for
searching for 1720~MHz OH masers associated with the shock interactions
of SNRs with molecular clouds.  In addition, it is an ideal candidate for
searching for conjugate OH satellite lines toward the nucleus due to the
large OH column densities and high radiation fields in this region.

\section{OBSERVATIONS AND DATA REDUCTION}
\label{sec-obs}
NGC~253 was observed in the two OH satellite lines at 1720.527~MHz and
1612.231~MHz on 17 November 1996, using the Very Large Array of the
National Radio Astronomy Observatory.\footnote{The National Radio
Astronomy Observatory is a facility of the National Science Foundation,
operated under a cooperative agreement by Associated Universities, Inc.}
The observations of NGC~253 were made in the highest resolution
configuration (A--array) of the VLA and provided an order of magnitude
improvement in the spatial resolution over the previous OH study of
NGC~253 (Turner 1985).  High resolution observations were desired to
search for the possible association of maser regions with discrete radio
sources within the complex nuclear region.

The phase reference used for the radio observations was the nuclear
position of $\alpha$(B1950)$=00^{\rm h} 45^{\rm m}05\fs7$;
$\delta$(B1950)$=-25\arcdeg 33\arcmin 40\arcsec$, and the observed
frequencies were centered on a LSR velocity of 200~km~s$^{-1}$, similar
to the systemic LSR velocity of 229~km~s$^{-1}$ derived from CO
observations (Canzian, Mundy, \& Scoville 1988).  We observed using a
3.125~MHz correlator bandwidth divided into 128 spectral channels,
resulting in a velocity resolution of 4.3~km s$^{-1}$and 4.5~km s$^{-1}$
at 1720~MHz and 1612~MHz respectively.  In addition to the line data, we
recorded the continuum data which represents the mean over the central
75\% of the bandwidth.  We observed only right handed circular
polarization at each frequency.  For primary flux and passband
calibration, we observed 0137+331 and 0542+498.  The radio source
0118--216 was used as the secondary phase and gain calibrator.

The data were calibrated according to standard VLA procedures using the
AIPS software package.  The continuum data were first carefully analyzed
to flag (u,v) visibilities showing interference.  Considerable editing
was required for the 1612 MHz data due to interference from the GLONASS
satellite system (Combrinck, West, \& Gaylard 1994).  After standard
passband, phase, and gain calibration of the continuum data, we improved
the data slightly by self--calibrating on the strong continuum of
NGC~253.  The line data were edited and calibrated by applying the
flagging commands and the calibration solutions derived for the
continuum data.

The continuum was subtracted from the line data in the (u,v) plane using
the AIPS task UVLIN\footnote{Words in capitals denote standard
procedures in AIPS.} which linearly interpolates the continuum across
the region showing line emission.  Since the 1720 and 1612~MHz line
features are weak compared to the strong continuum, several different
methods were used to test the reliability of the continuum subtraction.
In summary UVLIN produced results similar to those of IMLIN which
subtracts the continuum in the image plane.  For subtraction in the
(u,v) plane, the task UVLIN proved to be far superior to UVBAS in
removing artifacts from bright continuum sources, since UVLIN fits the
real and imaginary parts of the visibility data instead of fitting the
amplitudes and phases (Cornwell, Uson, \& Haddad 1992).

Continuum maps and continuum--free line cubes were made at both
frequencies using the AIPS task IMAGR.  For the strong continuum
emission, CLEAN algorithms were used, while for the line data only
``dirty'' cubes (no CLEAN components) were made.  We used both natural
and robust visibility weighting to produce the images.  Robust weighting
is a compromise between uniform and natural weighting (Briggs 1995)
which increases the spatial resolution at the expense of increasing the
noise level.  By adopting a robustness parameter of zero, we improved
the image resolution by approximately 40\% at the expense of increasing
the rms noise by 20\%.  The $1\sigma$ rms of the natural weighted
1720~MHz data is 1.2~mJy, while the 1612~MHz rms is 2.0~mJy, due to the
higher interference at 1612~MHz.  Interestingly, the rms noise at
1612~MHz does not degrade with robust weighting.  Upon further analysis
we found that the low level interference affects the data on the shorter
baselines to a larger degree which could account for the noise
characteristics of the 1612~MHz data with different (u,v) weighting.  By
using robust--weighting, we achieved a synthesized beam size of
$1\farcs7 \times 1\farcs0$ and $1\farcs8 \times 1\farcs1$ at 1720~MHz
and 1612~MHz respectively.
 
\section{RESULTS}

We show the 1720 and 1612~MHz natural--weighted channel maps in
Figures~1\&2.  The complexity of the OH emission and absorption features
in the nuclear region of NGC~253 is readily apparent.  The peak features have
narrow line widths and the limits on their brightness temperature of
T$_B > 1000$~K (T$_B > \frac{\lambda^2 S_{\nu}}{2k \Omega_{mb}}$) are
consistent with maser action.  The OH features follow the general sense
of the CO rotation curve (Canzian et al. 1988).  Since NGC~253 is viewed
nearly edge--on, the gas is spread over a broad range of velocities
(100--300~km~s$^{-1}$) at the nuclear position.  The position--velocity
distribution of the brightest OH features lie predominately along the
inner $x_2$ orbits of the bar potential described by Peng et al. (1996).

In order to improve our sensitivity to the extended weak structures seen
in Figure~1\&2, we have convolved the data to $3\farcs0 \times
3\farcs0$.  At the position of the brightest continuum emission
($\alpha$[B1950]$=00^{\rm h} 45^{\rm m}05\fs77$;
$\delta$[B1950]$=-25\arcdeg 33\arcmin 39\farcs7$), we find evidence for
conjugate emission and absorption in the 1720 and 1612~MHz lines
(Fig.~3).  To test for exact conjugate behavior, i.e., equal but
opposite line strengths, we smoothed the 1720~MHz data to match the
frequency resolution of 1612~MHz data cube.  We find conjugate behavior
over all velocities at the nuclear position of NGC~253 (Fig.~3).  As we
move off the nuclear position, we do not find this conjugate signature
over all velocities.  These results indicate a tendency for conjugate
behavior in the two satellite lines near the nucleus which may be due to
large column densities of OH and far--infrared pumping from the nuclear
activity, as seen in Cen~A (van Langevelde et al. 1995).

The individual spectral--line channel maps were inspected to search for
significant emission and absorption features.  In Table~1 we list the
ten OH features found to have peak emission or absorption line strengths
greater than five times the rms noise in the 1720 or 1612~MHz
robust--weighted channel maps.  The positions, peak fluxes, and
velocities of these features were determined by Gaussian fits to the
data.  All of the features were unresolved, suggesting angular source
sizes of $\la 1^{\prime\prime}$.  The reported errors are $1\sigma$, and
the upper flux density limits are $3\sigma$.

The locations of the ten detected OH features are shown in Figure~4
along with the positions of the compact radio sources given by UA97.
All ten of the features are located within the central 10 square
arcseconds of NGC~253 and are confined along the narrow ridge of the
brightest radio continuum emission.  Since the probability of finding a
$5\sigma$ point by chance in the central 10 square arcseconds of the
spectral--line cubes is only $0.02$, assuming Gaussian noise statistics,
all ten features are likely to be associated with real structures.  We
measure a position angle axis for the OH features of $48\pm4\arcdeg$
which is significantly smaller than the position angle of $64\arcdeg$
derived for the CO bar (Canzian et al. 1988) and the value of
$70\arcdeg$ found for the near--infrared bar (Forbes \& DePoy 1992).
However, the position angle of axis for the OH features is consistent
with that of the optical disk (Pence 1981) and that of the brightest
central CS and HCN emission regions (Peng et al. 1996; Paglione et
al. 1995).

In Figures~5--14 we plot the 1720 and 1612~MHz spectra at the position of
the ten features given in Table~1.  Many of the features appear as
narrow lines superimposed on a broader weak structure.  The broader
structure could represent the combination of many unresolved maser
regions.  If the broad structure is fitted and subtracted from the data,
the remaining narrow lines for features (5), (8), (9), and (10) are less
than $4\sigma$ in significance.  However, since features (9) and (10)
show structure at the same velocities in both the 1720 and 1612~MHz
data, we conclude that these features are real.  The 1720~MHz narrow
features of (5) and (8) do not show conclusive evidence for a
corresponding line at 1612~MHz, and it is possible that these features
are merely noise superimposed on the underlying broad structure.

Several of the features show evidence for opposite behavior in the 1720
and 1612~MHz lines, with one line in emission and the other in
absorption.  To test for conjugate line strengths, we convolved the
1720~MHz data to match the frequency and spatial resolutions of 1612~MHz
data cube.  The summation and difference of the 1720 and 1612~MHz data
at each position is shown in Figures~5--14.  Within the uncertainties of
the data, many of the features have conjugate line strengths.  We
discuss the individual features in detail in Sec.~4.3.

\section{DISCUSSION}
\label{sec-disc}
\subsection{Comparison with Previous OH Studies of NGC~253}

Early single--dish observations detected the presence of OH maser
activity in NGC~253 (Whiteoak \& Gardner 1973; Gardner \& Whiteoak
1975).  The single--dish data showed broad absorption in all four
ground--state transitions and narrow 1667~MHz emission features (Gardner
\& Whiteoak 1975).  Turner (1985) vastly improved upon the spatial
resolution of the single--dish work by mapping all four ground--state OH
lines with the VLA in C--array ($\sim
15^{\prime\prime}\times10^{\prime\prime}$) using low velocity resolution
(35 km~s$^{-1}$).  Most significantly, Turner found a nuclear outflow
plume extending 1.5 kpc above the equatorial plane in all of the OH
lines, except the 1720~MHz line.  Turner also detected broad central
absorption at all four transitions, but the 1612~MHz absorption was
weaker than the single--dish feature suggesting that the absorption is
extended and was partially resolved by the VLA.  In our higher
resolution data, we find even weaker central absorption, consistent with
extended gas being responsible for the majority of the absorption.

For detailed comparisons we have tapered the (u,v) range and convolved
the data to match the synthesized beam sizes for the data of Turner
(1985).  Unfortunately, we lack the sensitivity required to detect the
plume at 1612~MHz due to low level interference on short--baselines
(Sec.~\ref{sec-obs}).  We do not have these interference problems at
1720~MHz.  At 1720~MHz Turner (1985) report an emission region located
approximately $15\arcsec$ southwest of the nucleus at
140--175~km~s$^{-1}$.  We fail to confirm this feature in our data
convolved to matched the resolution of Turner's data.  Instead, we find
emission at similar velocities (Fig.~3) located at the nuclear position.
Since the strength of this emission ($\sim 10$ mJy) in our
high--resolution data is similar to that seen in the single--dish data
(Gardner \& Whiteoak 1975), we conclude that the bulk of this emission
arises from compact sources (i.e., consistent with many masers covering
the velocity range).

\subsection{Origin of the OH Features}
\label{sec-origin}
All of the ten detected OH features (Table~1) have narrow line widths,
are unresolved at arcsecond resolution, and have brightness temperatures
of T$_B > 1000$~K.  These properties suggest that maser action is the
likely origin for the OH features.  Assuming a maser origin, the
monochromatic luminosities of $(6-12)\times10^{4}$~Jy~kpc$^{2}$
($S\times D^{2}$) for the masers in NGC~253 are two to three orders of
magnitude more luminous than the observed Galactic satellite--line
masers (Gaume \& Mutel 1987; Frail et al. 1994; Yusef--Zadeh et
al. 1996).  This could be indicative of several components within the
20~pc projected synthesized beam diameter or could be due to the high
continuum level in NGC~253.  The high luminosities of the satellite
lines in NGC~253 may not be unusual for extragalactic systems.  In fact,
the lines in NGC~253 are relatively weak in comparison to those seen in
Cen~A which are a factor of 50--100 times more luminous (van Langevelde
et al. 1995).

As noted earlier, the locations of the OH emission and absorption
features all lie along the bright central ridge of the radio continuum
emission associated with SNRs and HII regions (UA97), as well as near
young stellar clusters (Watson et al. 1996).  Interestingly, the
distribution of the OH features are more narrowly confined than the
radio source distribution (Fig.~4).  Assuming that the radio continuum
emission and infrared radiation are well correlated, as expected in
star--forming galaxies (Condon 1992), these results may suggest that the
strong infrared radiation fields have a large influence on the existence
of the OH masers.  The OH features also appear to be coexistent, both
spatially and kinematically, with regions of dense molecular gas traced
by the HCO$^{+}$ (Carlstrom et al. 1990), HCN (Paglione et al. 1995),
and CS molecules (Peng et al. 1996).  The association of the OH features
with dense molecular gas and regions of star--formation leads to several
possible mechanisms for producing the masing activity in the central
nuclear regions of NGC~253.  Both collisional and infrared pumping
mechanisms are possible due to the high gas densities and strong
infrared fields.  We summarize the various satellite--line signatures
for the likely mechanisms producing the OH maser features below.

We expect at least three different mechanisms for producing OH maser
activity in the nuclear regions of NGC~253: (1) collisionally pumped
regions from SNR shocks, (2) radiatively pumped regions near sources of
strong far--infrared fields, and (3) combined mechanisms associated with
compact HII regions.  Method~(1) has been proposed to explain the
1720~MHz masers associated with Galactic SNRs interacting with molecular
clouds (Frail, Goss, \& Slysh 1994; Frail et al. 1996).  When the shock
wave from a SNR propagates into a molecular cloud, the gas is heated,
and the 1720~MHz line can be inverted from collisions of the OH
molecules with H$_2$.  Elitzur (1976) showed that the 1720~MHz OH line
is strongly inverted ($-\tau > 1$) by collisions for a range of
molecular gas densities of $10^{3}-10^{5}$~cm$^{-3}$ and temperatures of
T$\sim 20-200$~K, which is consistent with the conditions expected in
the post--shocked gas.  In addition, fast SNR shocks photodissociate the
H$_2$ and induce chemical reactions that enhance the OH abundance in the
post--shocked material (Hollenbach \& McKee 1989).  The studies of the
nuclear molecular gas in NGC~253 indicate warm gas and high gas
densities (Wall et al. 1991; Paglione et al. 1995) which are sufficient
for collisional inversion.  Since the high H$_2(\lambda =2.218 \micron$)
to CO ratios suggest the importance of shocked gas due to SNRs in the
central regions of NGC~253 (Prada et al. 1996), we could expect a
significant fraction of the OH molecules to be collisionally inverted.
For the collisionally pumped scenario at temperatures $T < 200$~K, we
would observe a strong 1720~MHz line in emission, and no associated
1612~MHz feature.

Radiative pumps can produce strong satellite line masers, as seen for
example in the 1612~MHz masers associated with OH/IR stars (e.g., Booth
et al. 1981).  In conditions where the 1720 and 1612~MHz lines compete
for the same infrared pumping photons associated with the rotational
transitions between the ground state ($^{2}\Pi_{3/2}\,J=3/2$) and the
first excited levels of $^{2}\Pi_{3/2}\,J=5/2$ ($119\micron$) and
$^{2}\Pi_{1/2}\,J=1/2$ ($79\micron$), the 1720 and 1612~MHz lines show
conjugate behavior, with one line showing stimulated emission and the
other showing stimulated absorption (Elitzur 1992).  Such behavior has
been observed in the nuclear regions of NGC~4945 (Whiteoak \& Gardner
1975), and more recently in Cen~A (van Langevelde et al. 1995) and M82
(Seaquist et al. 1997).  When the OH molecules are predominately pumped
by the rotational intraladder transition at 119$\micron$, the 1720~MHz
line shows stimulated emission while stimulated absorption is seen in
the 1612~MHz line.  The reverse occurs if the cross--ladder rotational
transition at 79$\micron$ provides significant pumping.  The dominant
pumping transition depends on the OH column density.  For OH column
densities of $N({\rm OH})/\Delta V \simeq 10^{14}$
cm$^{-2}$~km$^{-1}$~s, the 119$\micron$ transition becomes optically
thick, and the 1720~MHz line is in emission while the 1612~MHz line is
in absorption.  At OH column densities higher than approximately $N({\rm
OH})/\Delta V \simeq 10^{15}$ cm$^{-2}$~km$^{-1}$~s, the 79$\micron$
transition becomes optically thick producing 1612~MHz in emission and
1720~MHz in absorption (van Langevelde et al. 1995).  Assuming
line--widths of order 1~km~s$^{-1}$ and typical OH/H$_2$ abundance
ratios of order $10^{-7}$ (Langer \& Graedel 1989), the transitional OH
column density of $10^{15}$cm$^{-2}$~km$^{-1}$~s corresponds to a H$_2$
column density of $10^{22}$~cm$^{-2}$.  This column density is similar
to the average H$_2$ column density of $3 \times 10^{22}$~cm$^{-2}$
implied by the CO data for NGC~253 (Canzian et al. 1988) and is a
typical value for active galaxies.  This is an interesting result.
There is no reason to expect, a~priori, the transitional OH column
density to correspond with the column densities found in active
galaxies.  It is possible that the conjugate behavior seen previously in
Cen~A (van Langevelde et al. 1995) and found here for NGC~253 (Fig.~3)
is common for the molecular--rich centers of most galaxies that have
strong infrared fields.

Since the convolved data at the nuclear position clearly shows conjugate
behavior (Fig.~3), the variation in the line strengths could be
interpreted as a variation in the molecular column density.  In this
scenario, the low velocity gas (100--180~km~s$^{-1}$) showing 1720~MHz
emission would represent gas with OH column densities less than the
transitional value of $10^{15}$cm$^{-2}$~km$^{-1}$~s, while the gas at
velocities of 230--300~km~s$^{-1}$ showing 1612~MHz emission would have
OH column densities higher than the transitional value.  The broad 1720
and 1612~MHz emission features in Figure~3 are consistent with the
velocities of the two bright features in the CS position--velocity
diagram (Fig.~6 of Peng et al. 1996).  Although there is qualitative
agreement, the CS data do not show an increase in the molecular column
density as a function of velocity at the nuclear position, as suggested
by the simple interpretation of the OH data.  Perhaps, these results
indicate variations of the CS/OH abundance ratio in NGC~253.

A third proposed origin for producing the OH features in NGC~253 is
associated with compact HII regions.  Until this decade the observations
have been way ahead of the theoretical interpretation of OH masers
associated with HII regions.  However, detailed theoretical work in the
last ten years has vastly improved our understanding of the important
physical processes (e.g., collisions, infrared radiation, local line
overlap, nonlocal line overlap, and accelerated and decelerated flows)
involved in star--forming regions (Cesaroni \& Walmsley 1991; Gray,
Doel, \& Field 1991; Gray, Field, \& Doel 1992; Pavlakis \& Kylafis
1996a; Pavlakis \& Kylafis 1996b).  Although early work by Elitzur
(1976) suggested that radiative pumps could not produce strong 1720~MHz
masers, the more recent work including the effects of infrared line
overlap contradict these findings.  The combination of radiative pumps
with other processes, such as line overlap, can at least qualitatively
explain many of the observed OH features seen in HII regions (e.g.,
Cesaroni \& Walmsley 1991).  Numerous OH maser transitions have been
detected in Galactic HII regions (Gaume \& Mutel 1987; Reid \& Moran
1981 and references therein), showing a variety of different line
ratios.  Under LTE with low OH optical depths, we would expect line
ratios of 1:1:5:9 for the ground state OH transitions of 1612, 1720,
1665, 1667~MHz, respectively.  These ratios are rarely seen.  For HII
regions the 1665~MHz is typically stronger by approximately an order of
magnitude over the other three ground state transitions (Reid \& Moran
1981).

\subsection{Individual OH features}
 
In this subsection we discuss the correspondence of the ten individual
compact OH features with discrete radio sources listed in UA97 and the
implications for the origin of their maser action.  Feature~(1) is near
the radio source designated 5.45-42.8 (UA97).  It is seen in absorption
against the background continuum at 1612~MHz and shares a velocity
component with nearby feature~(2) showing emission at 1720~MHz.
Feature~(2) is coincident with the second strongest radio source in
NGC~253, 5.48-43.3, which is thought to be one or more SNRs with a radio
luminosity of 20 times that of Cas~A (UA97).  The low--velocity side of
feature~(2) shows conjugate behavior in the satellite lines.  The
high--velocity side shows excess emission at 1720~MHz, as seen in the
summation of the 1612 and 1720~MHz data (Fig.~6).  Although somewhat
speculative, these results could suggest that the high--velocity
component is due to 1720~MHz inversion via SNR shock induced collisional
pumping and that the low--velocity component arises from infrared
pumping which would be consistent with the conjugate behavior seen in
the satellite lines (Sec.~\ref{sec-origin}).

Feature~(3) is positionally coincident with the compact radio source
5.54-42.2 which is likely an HII region due to its flat spectral index
(UA97).  The velocity and position of feature~(3) also agrees with that
of HCN complex~\#7 (Paglione et al. 1995) and CS complex~C (Peng et
al. 1996).  Both the 1612~MHz and 1720~MHz lines are seen at similar
levels of emission for feature~(3) which is consistent with the
observations of some Galactic OH/HII regions (Gaume \& Mutel 1987).  If
feature~(3) is associated with an HII region, we could expect OH
main--line emission from this source.  Although there is no evidence for
narrow 1665~MHz emission, the single--dish data do show a narrow
1667~MHz emission feature at the correct velocity of approximately
260~km~s$^{-1}$ (Gardner \& Whiteoak 1975).  Further VLA A--array
observations at 1667~MHz are required to test whether this emission is
located at the position of feature~(3).

The remaining features (4)--(10) are associated with the molecular HCN
regions~\#5\&6 (Paglione et al. 1995) and CS complex~B (Peng et
al. 1996).  Feature~(4) is located near the bright HII complex
5.72-40.1, which is similar to the central regions of 30~Doradus (UA97).
It is detected only in emission at 1720~MHz implying that collisional
pumping is a possible mechanism for the inversion.  Since the average
molecular conditions in NGC~253 are similar to those required for
collisional inversion, this feature could merely represent a region of
slightly enhanced temperatures and densities from weak shocks or bulk
motions associated with the local star--formation activity, not
necessarily strong SNR shocks.

Feature~(6) is coincident with the radio source 5.76-39.7.  It shows
strong emission at 1612~MHz and evidence for weaker 1720~MHz absorption.
These line characteristics are consistent with radiative pumping and
conjugate OH optical depths of $\tau \ga 1$, similar to that seen in the
strong OH maser source detected in M82 (Seaquist et al. 1997).  Since
the 1612~MHz emission is stronger than the corresponding 1720~MHz
absorption, we can derive the OH optical depth.  Assuming conjugate OH
optical depths ($\tau_{1720} = -\tau_{1612} =\tau$), the ratio of the
observed line intensities is $I_{1720}/I_{1612} = (\exp^{-\tau} -
1)/(\exp^{\tau}- 1)$.  The line intensity ratio of $-0.37$ for
feature~(6) implies an optical depth of $\tau \simeq 1$.  This OH
optical depth suggests a OH column density of $N({\rm OH})/\Delta V
\approx {\rm few} \times 10^{15}$ cm$^{-2}$~km$^{-1}$~s.  Since the
narrow peak of feature (6) has a line width of approximately
10~km~s$^{-1}$ and assuming $N({\rm OH})/N({\rm H}_2) \approx 10^{-7}$,
we estimate that $N({\rm H}_2) \approx {\rm few} \times
10^{23}$~cm$^{-2}$.  By using the limit on the radio source size of $D <
1$~pc for 5.76-39.7 (UA97), we derive a density of $n({\rm H}_2) >
10^{5}$~cm$^{-3}$.  This high molecular gas density is consistent with
the association of feature~(6) with the HCN and CS emission regions.
The velocity of feature~(6) is also very close to the velocity of the
strongest 1667~MHz line seen in the single--dish data (Gardner \&
Whiteoak 1975).  As for feature~(3), observations with the VLA in
A--array at 1667~MHz would confirm or refute the association of the
1667~MHz emission regions with the satellite--line sources detected
here.

The data for features~(5), (7), (9), \& (10) are consistent with
conjugate line strengths within the uncertainties of the data.
Feature~(5) is near 5.75-39.9 which has an unknown radio spectral index.
The position of feature~(7) is coincident with 5.78-39.4 which has a
steep spectral index, implying that one or more SNRs is the origin for
the majority of the radio emission.  Feature~(7) is also very near the
suspected AGN of NGC~253 (5.79-39.0, UA97; TH2, Turner \& Ho 1985).
Feature~(9) is coincident with 5.85-38.7, which has an unknown spectral
index, and feature~(10) is not coincident with any strong compact radio
source.  The fact that all of these sources appear to show conjugate
line strengths suggests this behavior is pervasive throughout the
central regions of NGC~253.

Feature~(8) has a broad underlying component with conjugate behavior in
the satellite lines, but the peak shows excess 1720~MHz emission.  Since
feature~(8) is located among a collection of SNRs, it is possible that
the excess 1720~MHz emission peak is due to collisional inversion from
the SNR shocks.  However, this interpretation is questionable since
the feature could merely represent a $2\sigma$ noise channel on top of
the broader conjugate feature (Sec.~\ref{sec-obs}).

Although many of the details in the above discussion are fairly
speculative, it appears that the OH features detected in NGC~253
represent several different types of sources.  We find examples of
sources consistent with 1720~MHz inversion associated with SNRs, as seen
for Galactic SNRs (Frail et al. 1996).  We also find examples of OH
maser activity associated with HII regions.  However, most of the
detected sources cannot simply be categorized as SNRs or HII regions.
Given that the luminosities of the OH features are two to three orders
of magnitude more luminous than their Galactic counterparts, the OH
maser sources detected here could represent a collection of ten or more
unresolved sources.  The broad low--level emission and absorption seen
throughout the central regions may also be due to many maser regions.
These observations appear to show a prevailing tendency for conjugate
behavior in the satellite--lines which is likely due to pumping from the
strong infrared radiation fields associated the star--formation activity
in NGC~253.  It is also interesting to note that the regions showing
excess emission at 1612~MHz, which implies sources with the largest OH
column densities, are located nearest to the AGN (seen at $\sim
220-300$~km~s$^{-1}$ in the spectra of features [6], [7], \& [8]).

\section{Conclusions}

We have detected several compact 1720~MHz and 1612~MHz OH features in
the nuclear region of NGC~253.  These features are likely OH maser
regions associated with the star--formation activity in NGC~253.  The OH
features appear physically associated with the dense molecular gas and
several compact SNRs and HII regions.  Although we find evidence for
1720~MHz maser activity consistent with collisional inversion due to the
interaction of SNRs with the molecular gas, as proposed for Galactic SNRs
(Frail et al. 1996), these regions are not the dominant source of OH
satellite--line masers in NGC~253.  Instead, we find conjugate behavior
in the satellite lines that is pervasive throughout the nuclear regions
of NGC~253.  These results suggest that radiative pumping is the
dominant mechanism for inverting the OH molecules.  The conjugate
signature may be common for the centers of most active galaxies, since
this behavior probes the appropriate temperature, density, and molecular
column density regimes typically found in active galaxies.

This research was supported by a grant to E.R.S. from the Natural
Sciences and Engineering Research Council of Canada.

\clearpage 
\figcaption[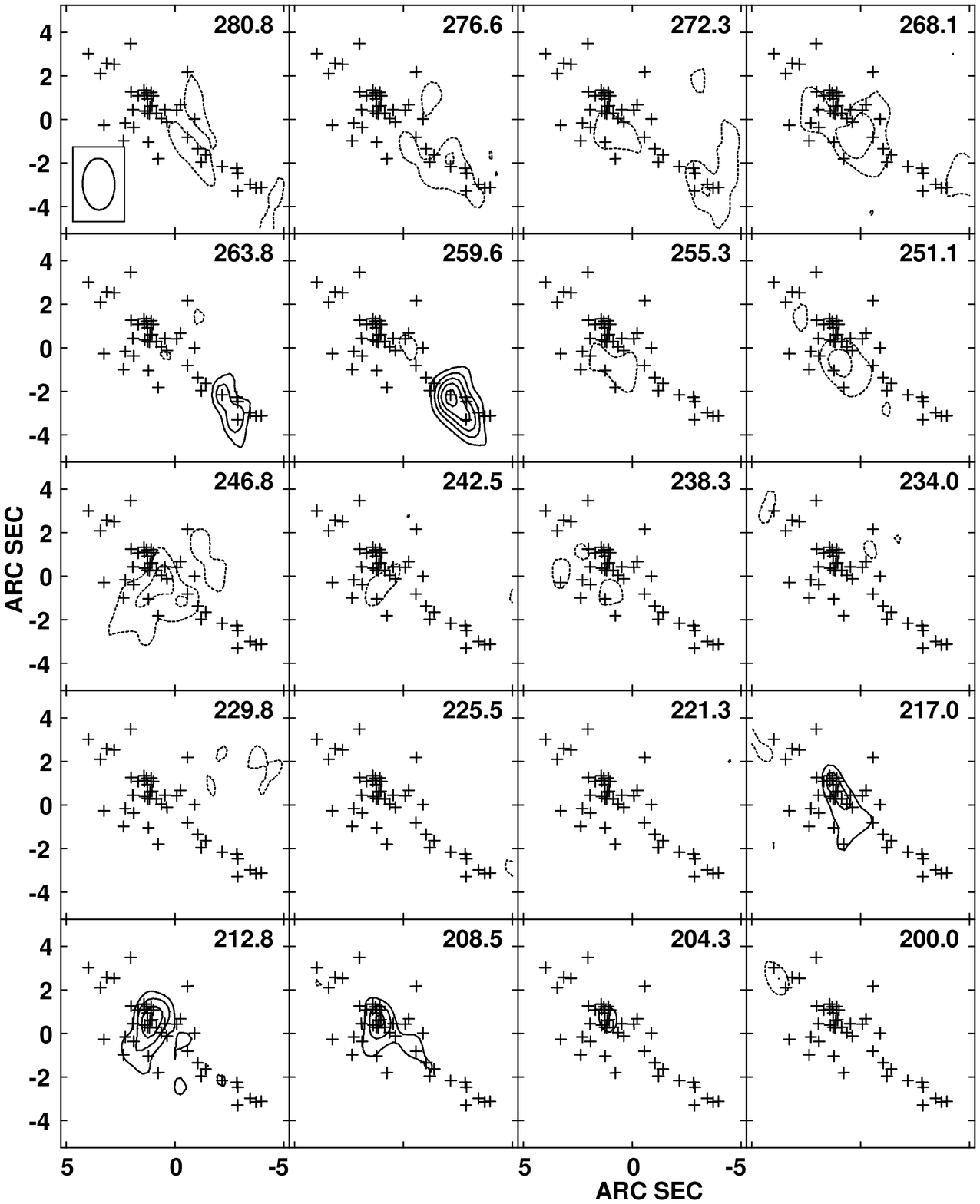,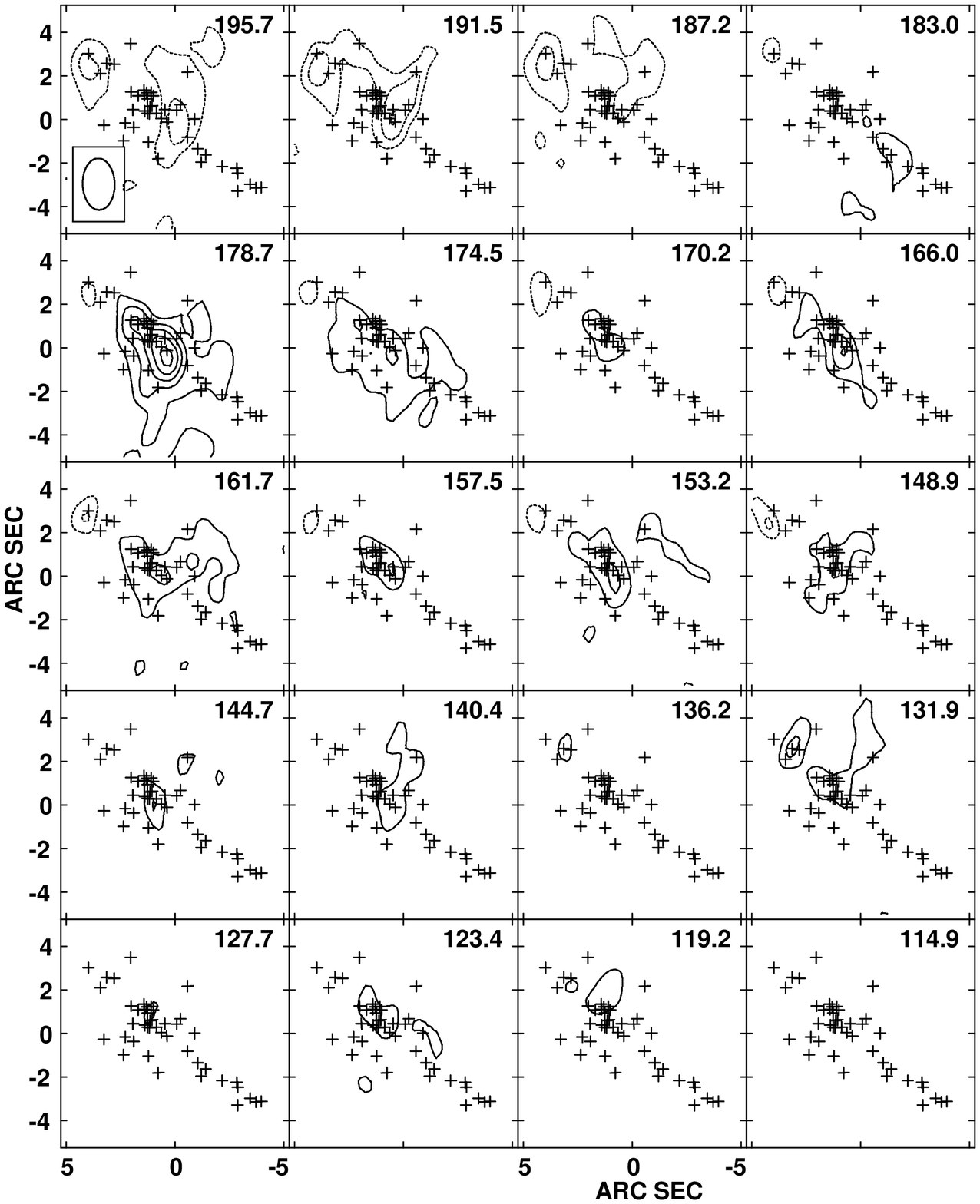]{The natural--weighted
channel maps for the 1720~MHz OH satellite line in NGC~253.  The radio
LSR velocity in km~s$^{-1}$ is given in upper--right corner of each
panel, and the synthesized beam is shown in the upper--left panel.  The
contour levels are 1~mJy/beam$\times (-12, -10, -8, -6, -4, 4, 6, 8, 10,
12)$, and the 1$\sigma$ rms is 1.2~mJy/beam.}

\figcaption[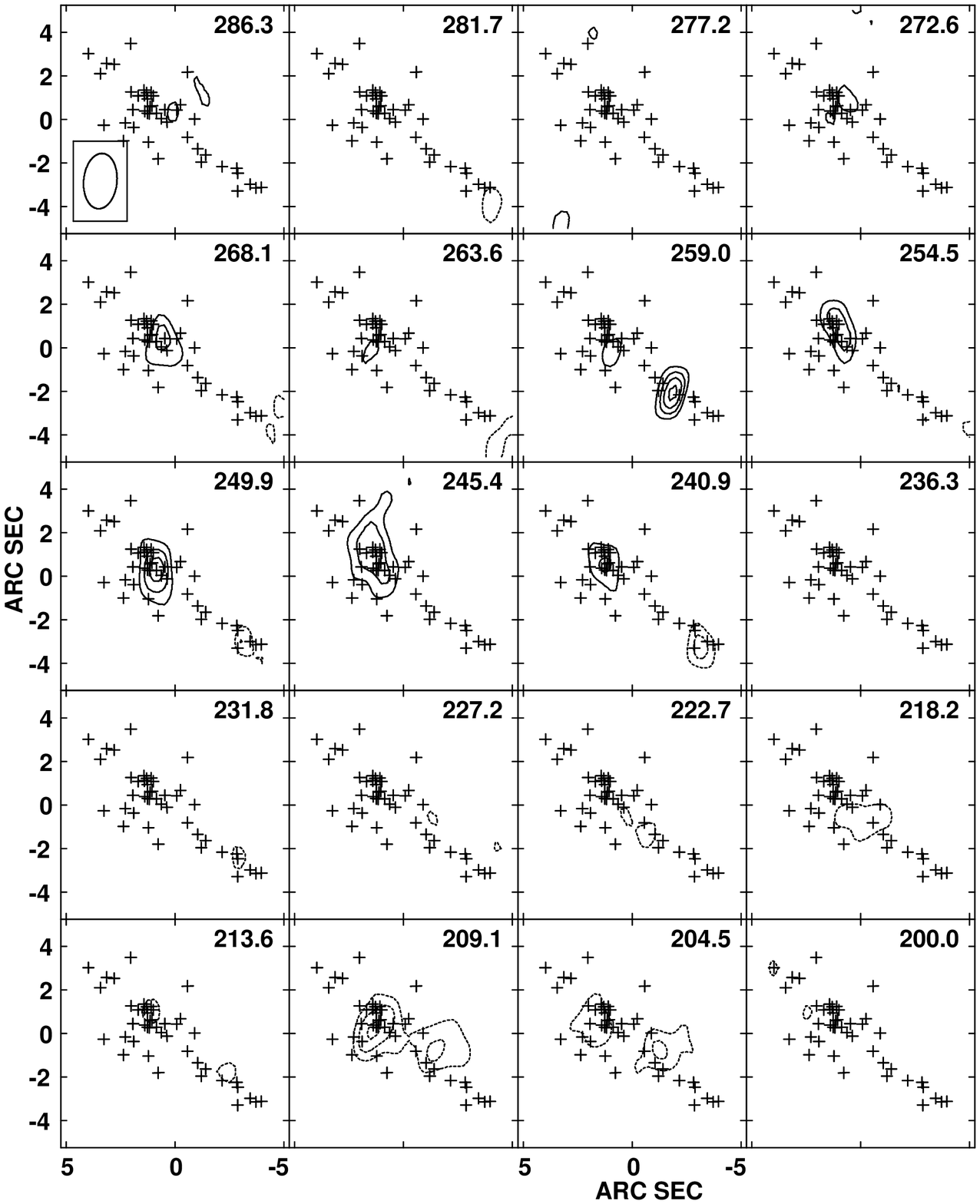,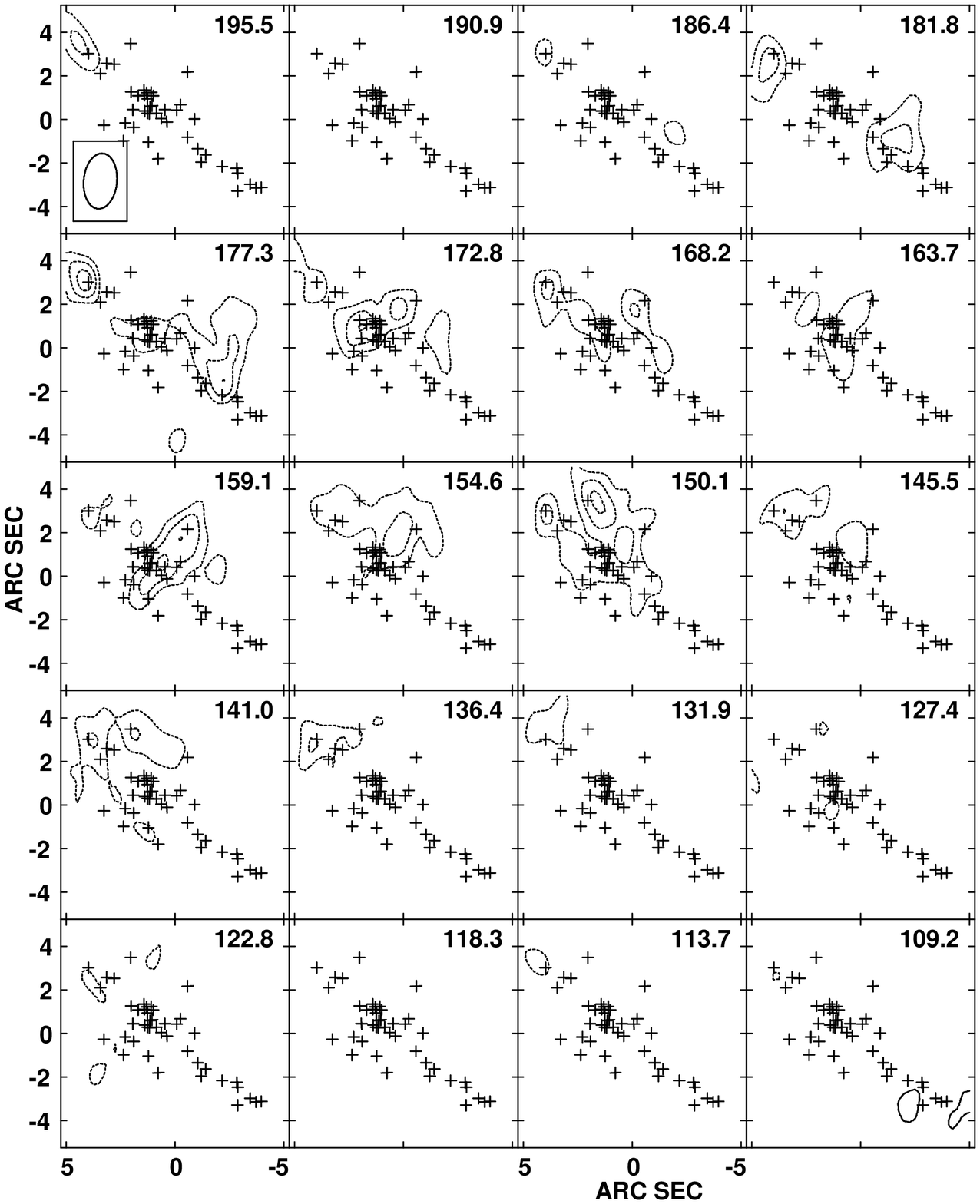]{The natural--weighted
channel maps for the 1612~MHz OH satellite line in NGC~253.  The contour
levels are 1~mJy/beam$\times (-12,-10,-8,-6,6,8,10,12)$, and the
1$\sigma$ rms 2.0~mJy/beam. As in Figure~1, the radio LSR velocity in
km~s$^{-1}$ is given in upper--right corner of each panel, and the
synthesized beam is shown in the upper--left panel.}

\figcaption[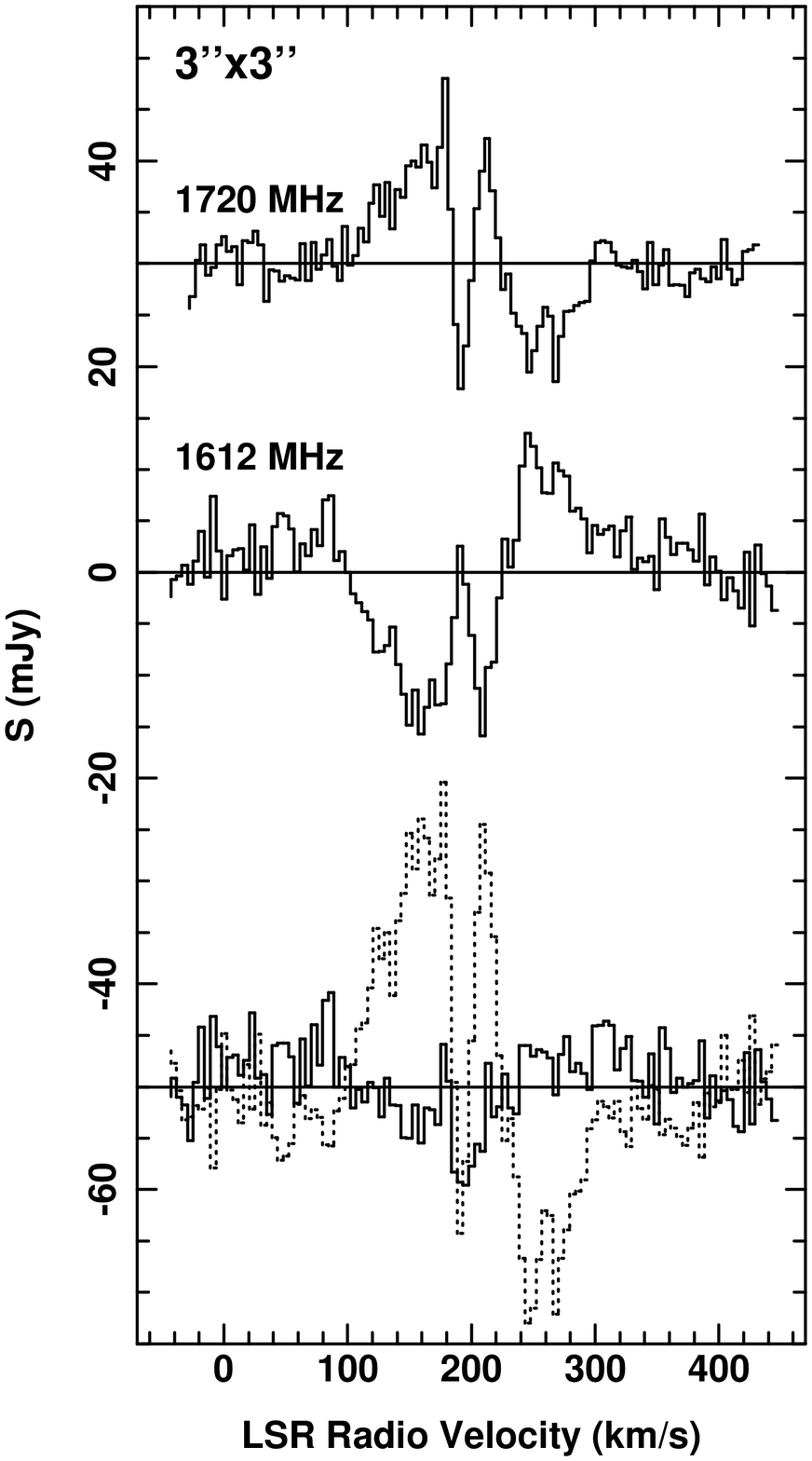]{The spectra at the position of the
brightest continuum emission ($\alpha$[B1950]$=00^{\rm h} 45^{\rm
m}05\fs77$; $\delta$[B1950]$=-25\arcdeg 33\arcmin 39\farcs7$) for the
convolved ($3\farcs0\times3\farcs0$) natural--weighted data.  In the
lower spectra, the solid line is the sum of the 1720~MHz (upper
spectrum) and 1612~MHz (middle spectrum) data, while the dotted line
represents their difference.  These results demonstrate the conjugate
behavior of the satellite lines near the nucleus of NGC~253.}

\figcaption[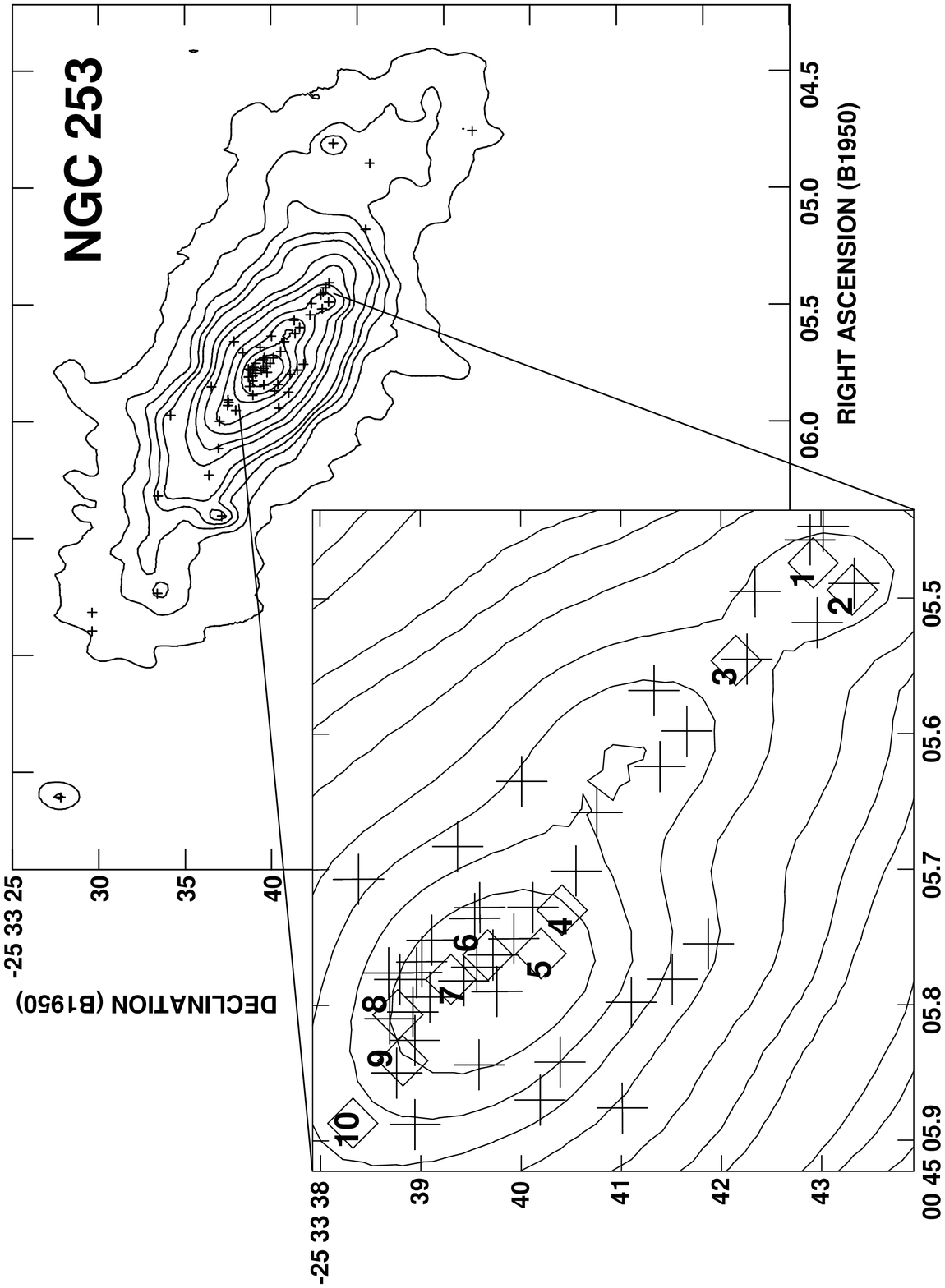]{The 1720~MHz continuum map of NGC~253.  The
positions of discrete radio sources from Ulvestad \& Antonucci (1997)
are given as cross symbols, and the locations of the candidate maser
regions (Table~1) are shown as diamond symbols in the magnified image of
the nucleus.  The contour levels are are 1~mJy/beam$\times$(1, 3, 5, 7,
9, 15, 20, 30, 50, 70, 90, 110).}

\figcaption[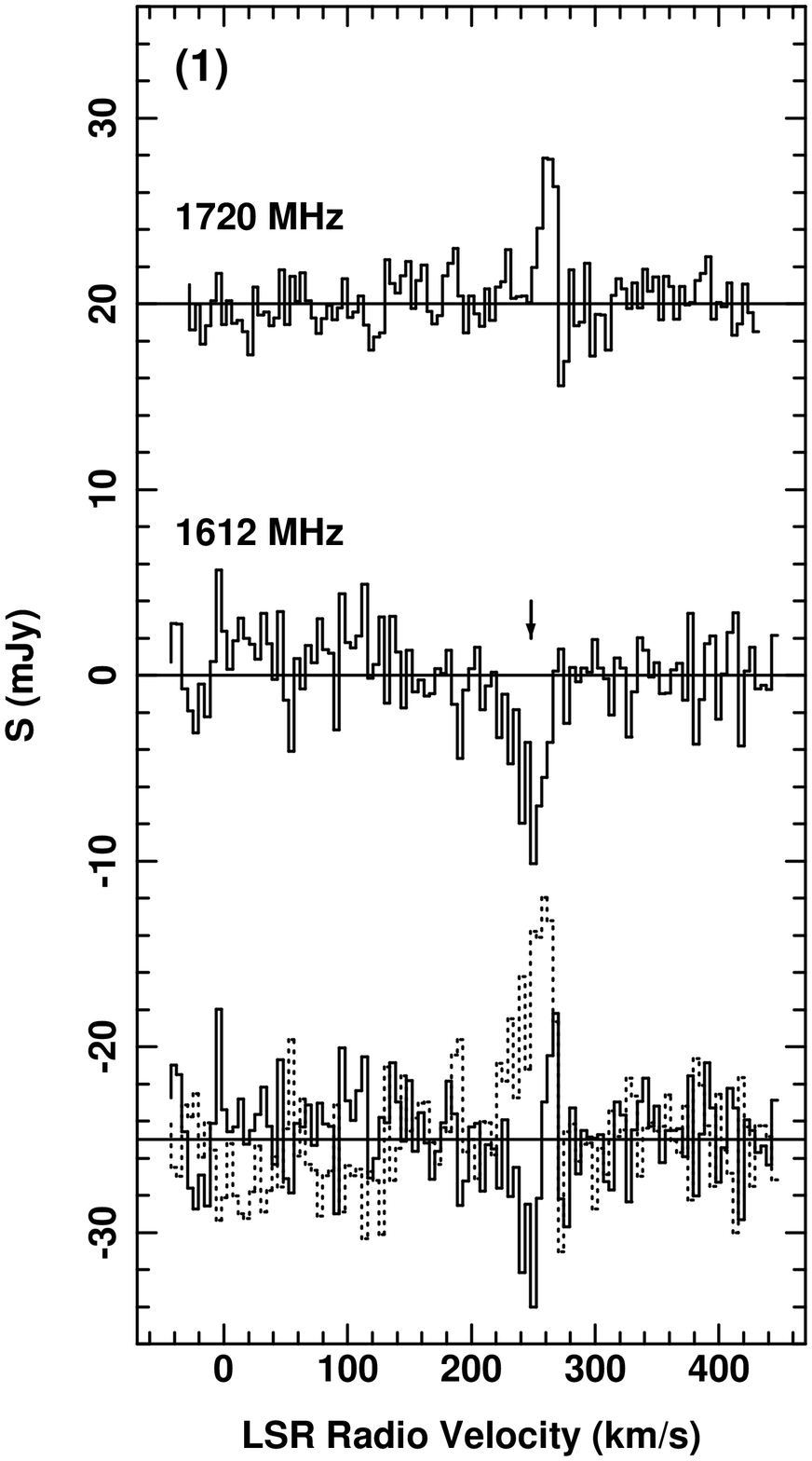]{The 1720~MHz (upper spectrum) and 1612~MHz
(middle spectrum) spectra at the position for feature (1) listed in
Table~1.  The velocity of this feature is marked by an arrow.  The solid
line in the lower spectra represents the sum of the 1720 and 1612~MHz
data, while the dotted line represents their difference.}

\figcaption[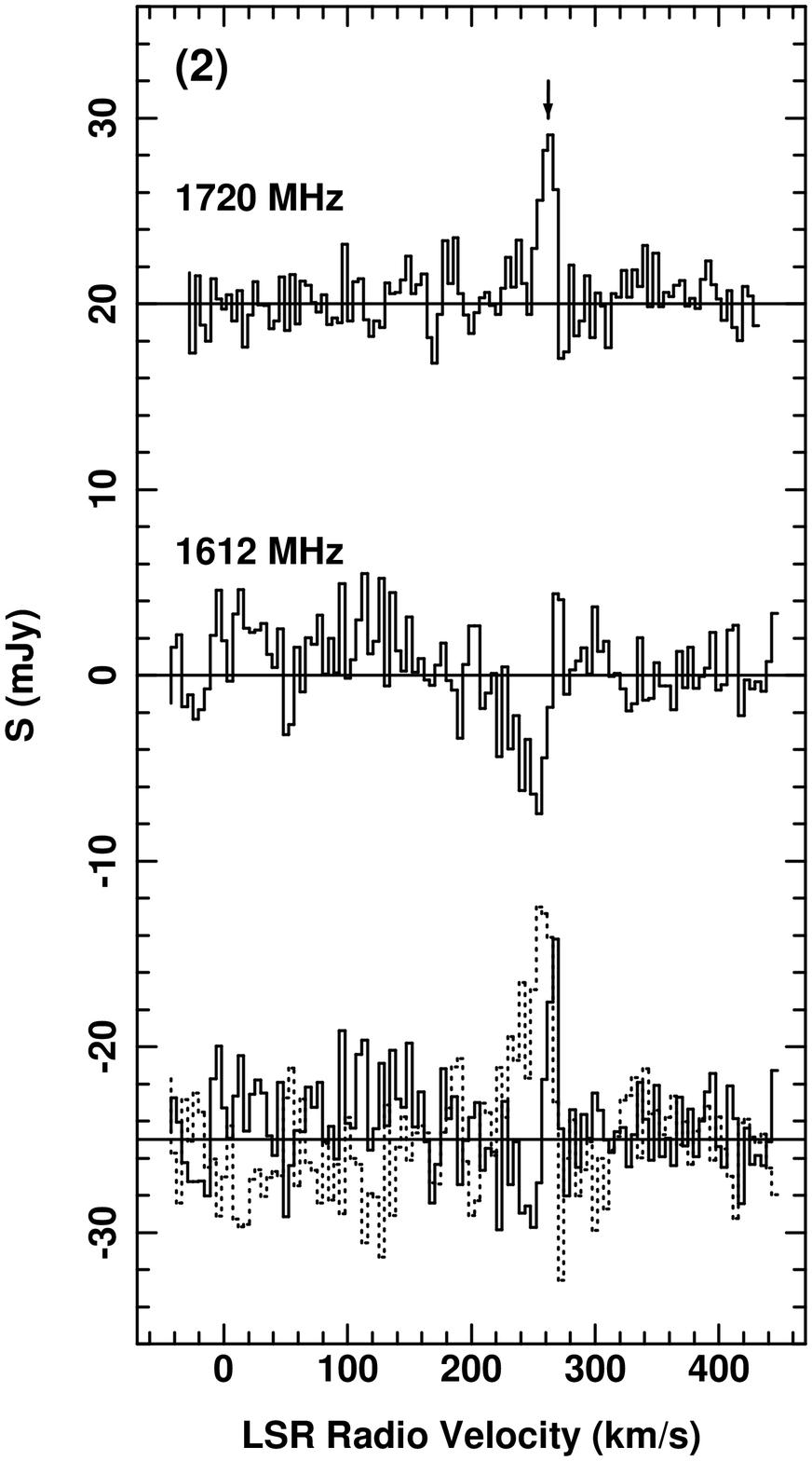]{The spectra for feature (2) as described in
Figure~5.}

\figcaption[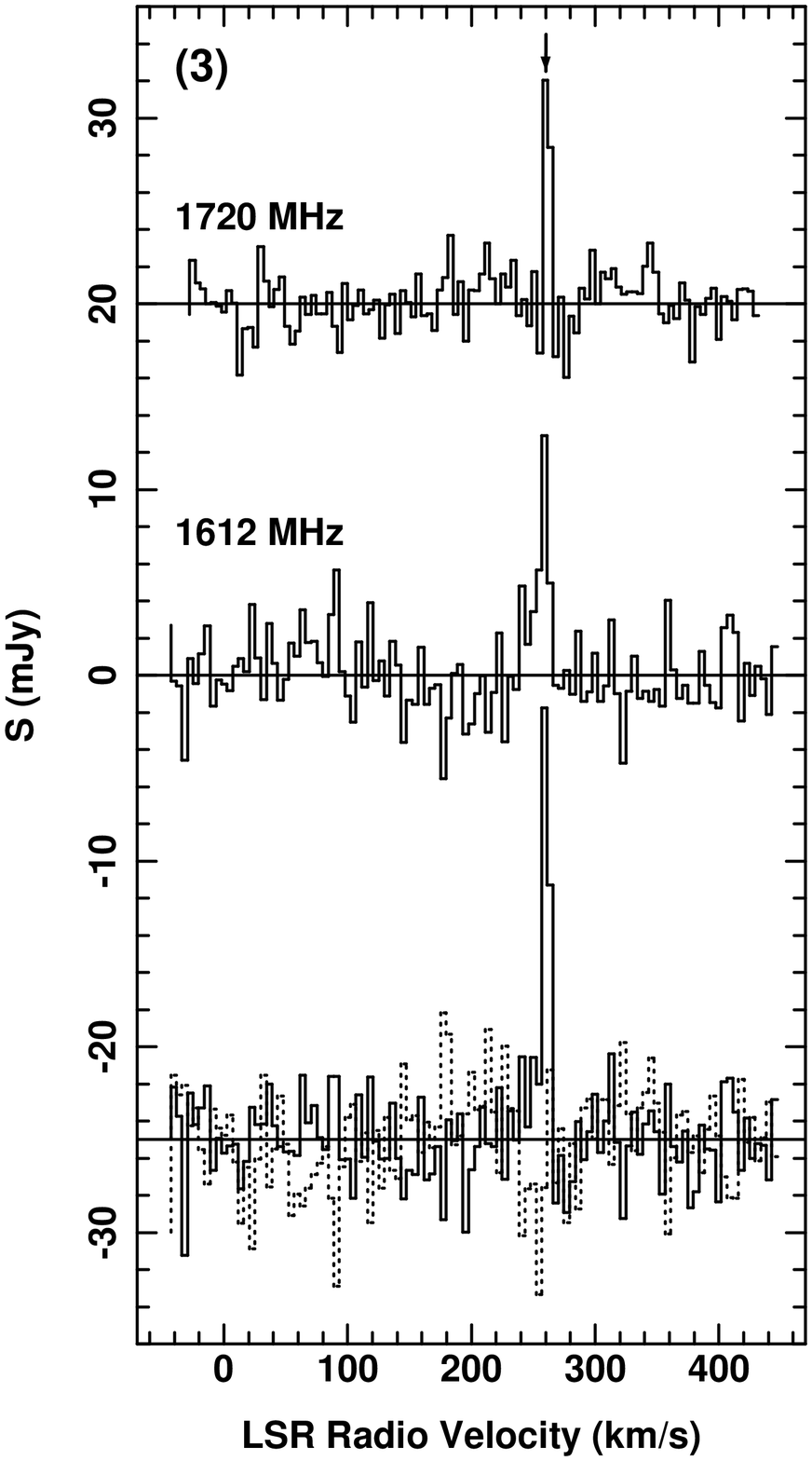]{The spectra for feature (3) as described in
Figure~5.}

\figcaption[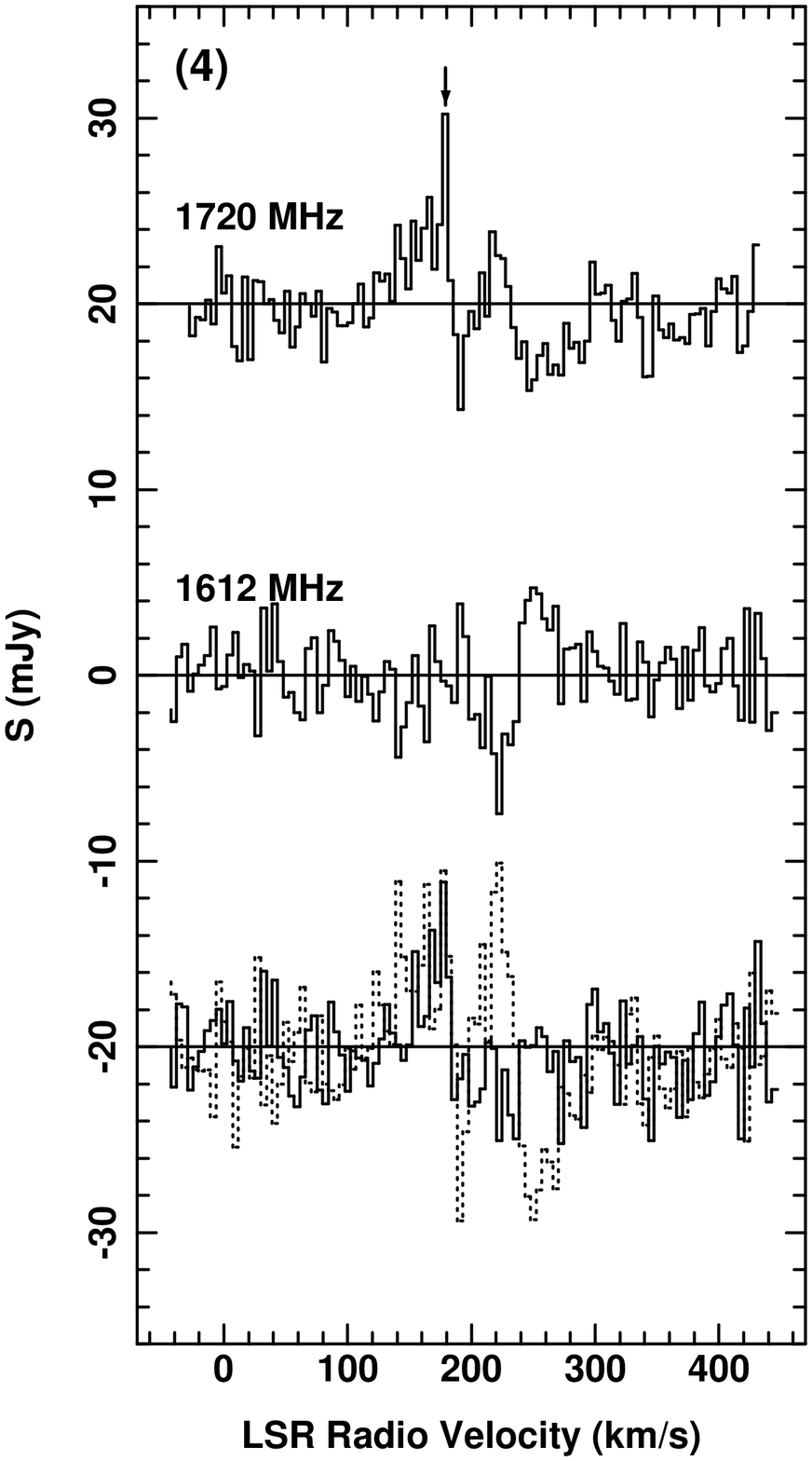]{The spectra for feature (4) as described in
Figure~5.}

\figcaption[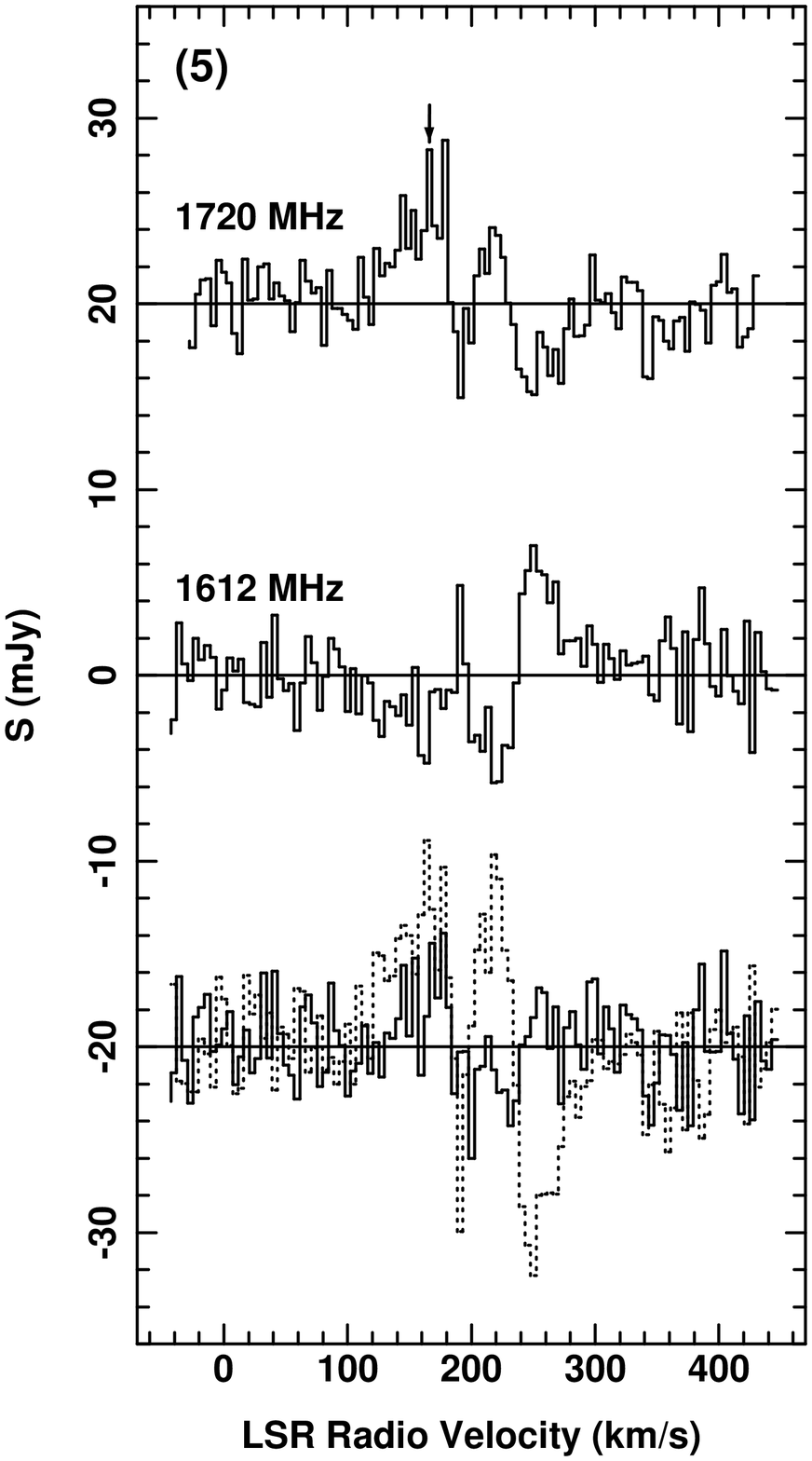]{The spectra for feature (5) as described in
Figure~5.}

\figcaption[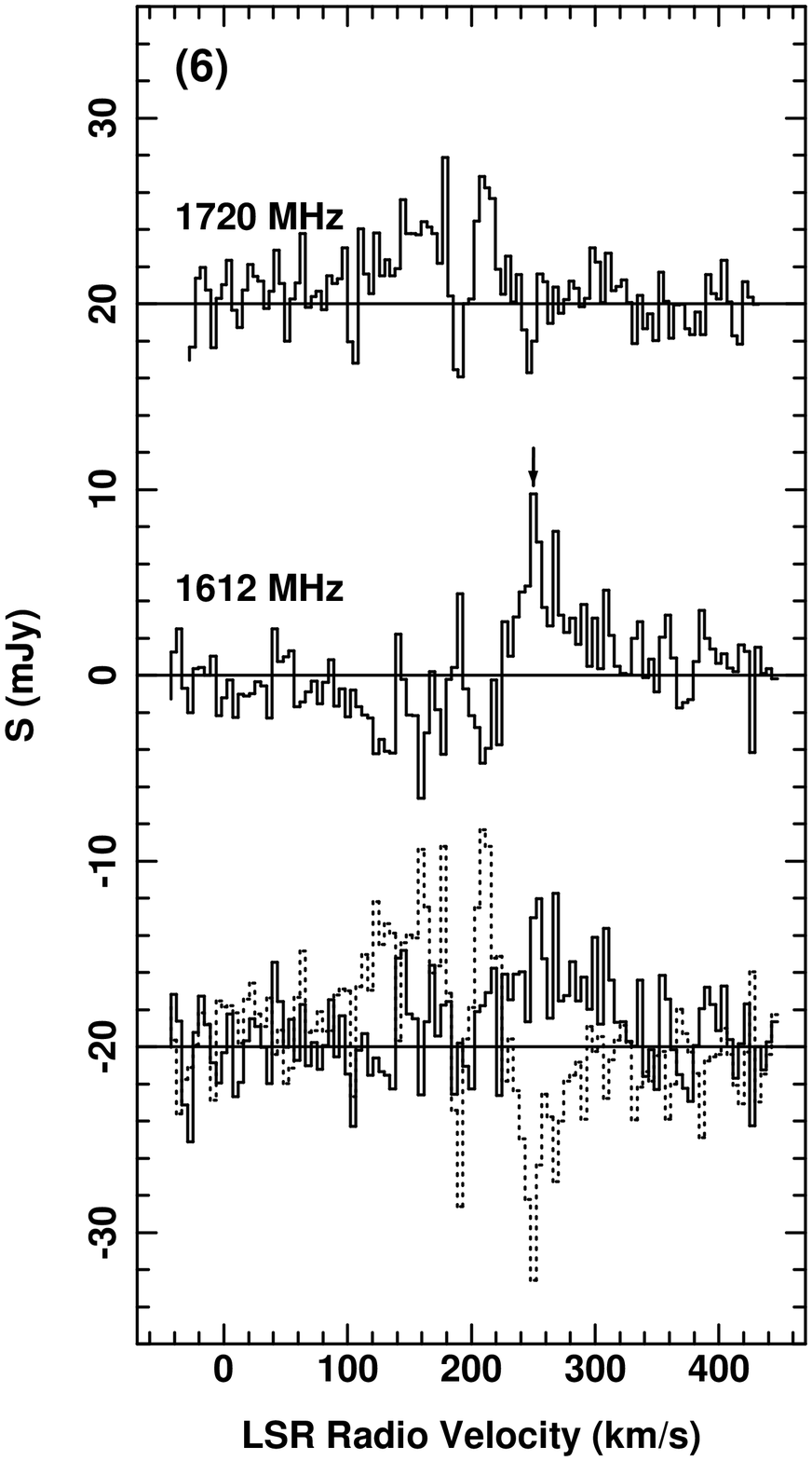]{The spectra for feature (6) as described in
Figure~5.}

\figcaption[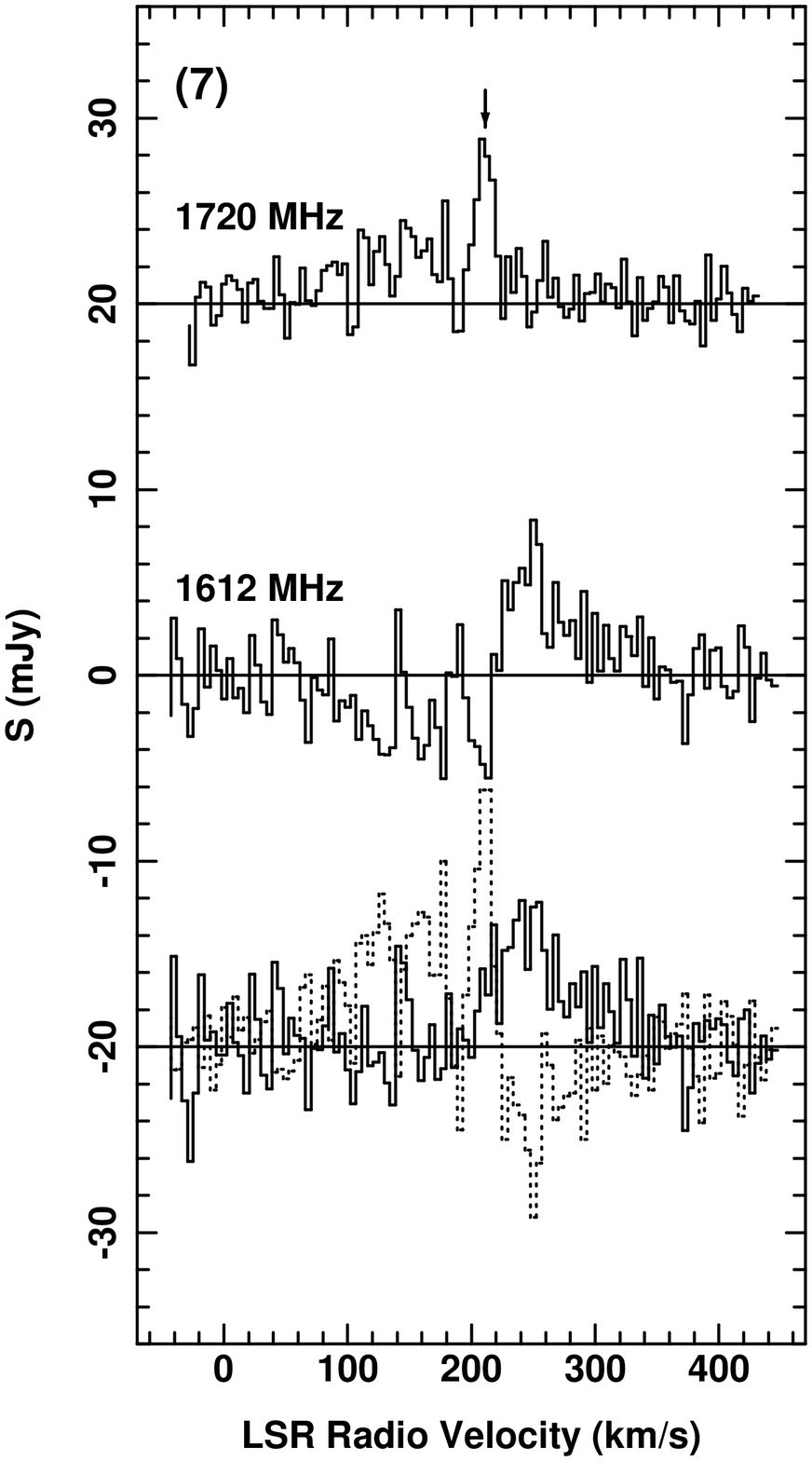]{The spectra for feature (7) as described in
Figure~5.}

\figcaption[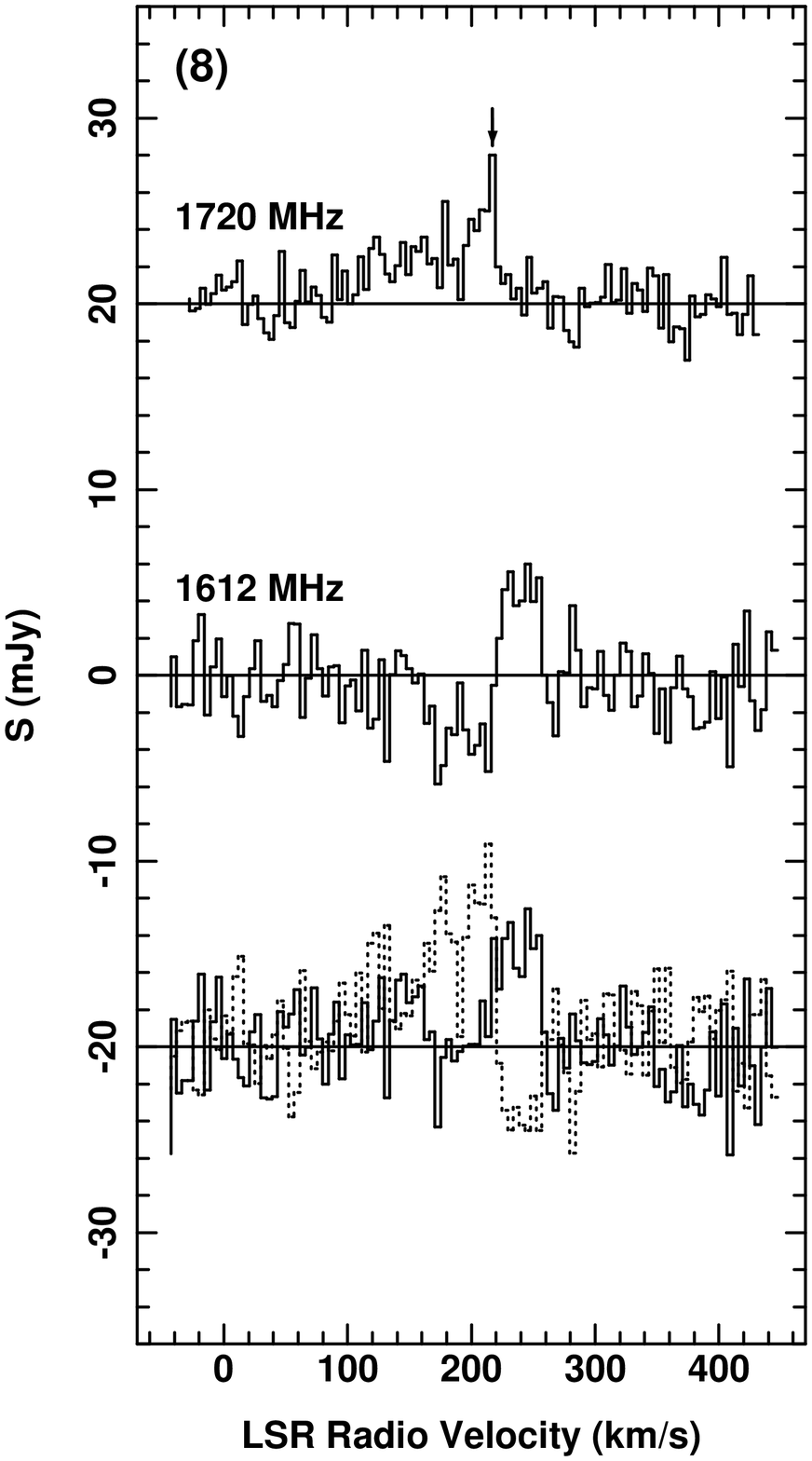]{The spectra for feature (8) as described in
Figure~5.}

\figcaption[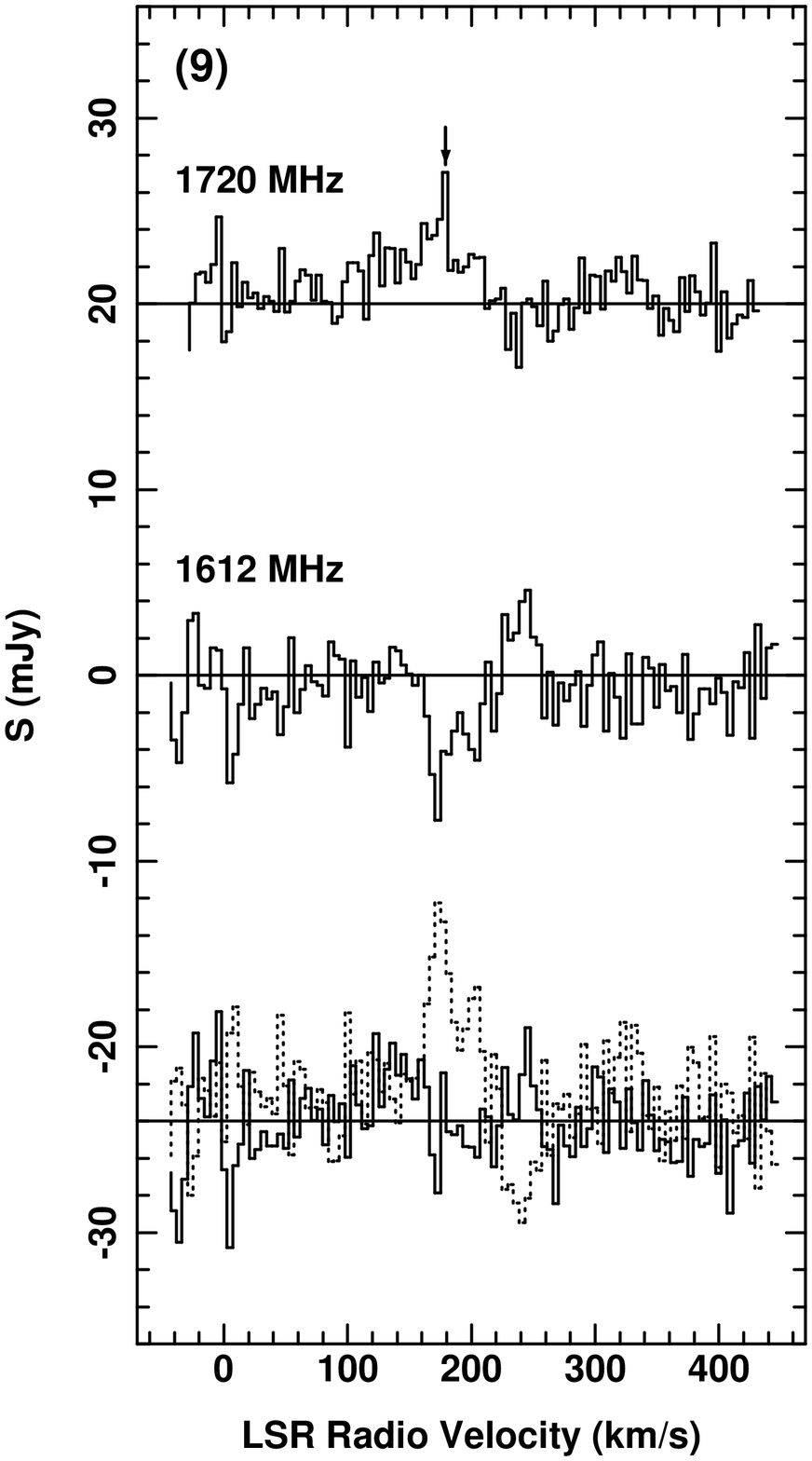]{The spectra for feature (9) as described in
Figure~5.}

\figcaption[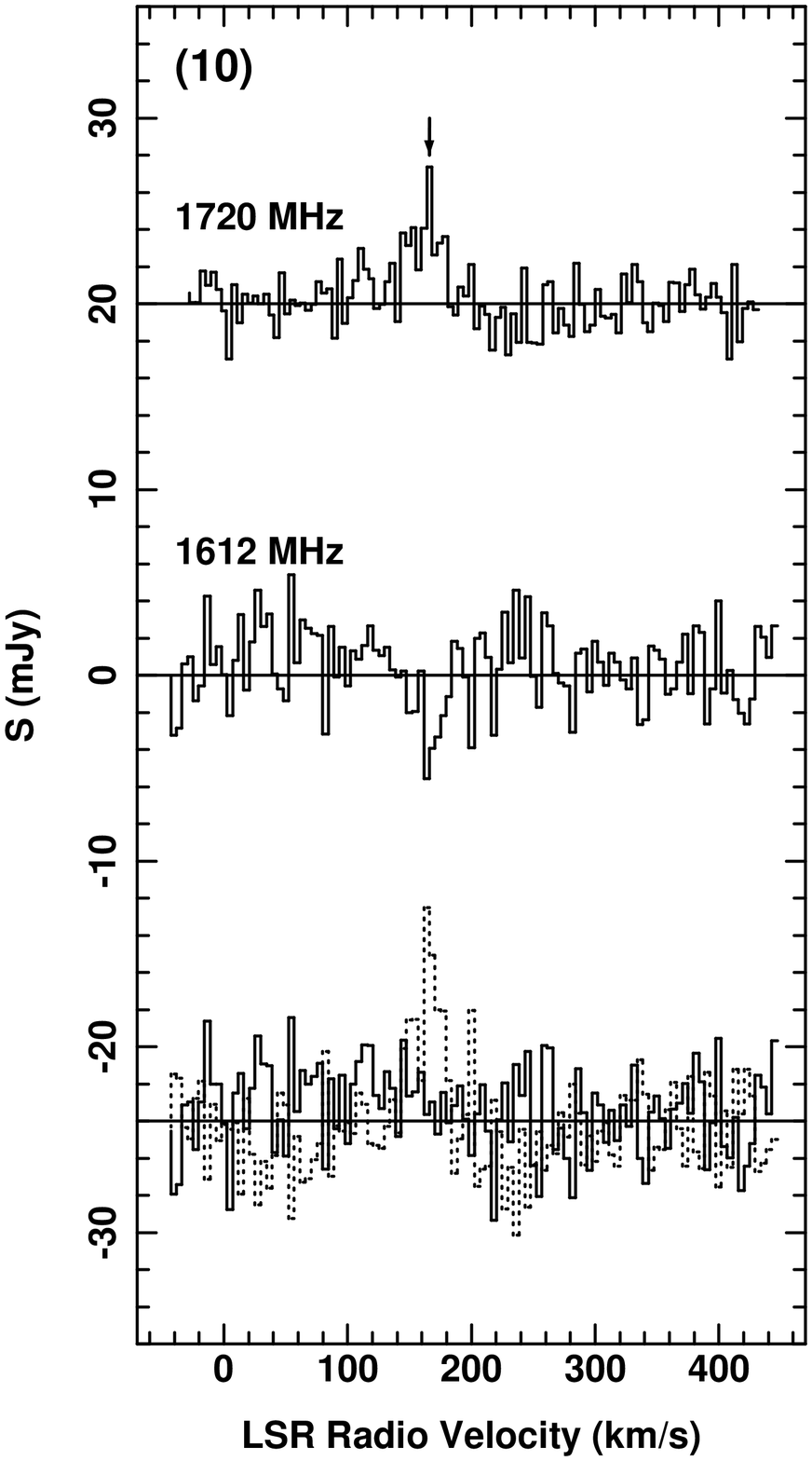]{The spectra for feature (10) as described in
Figure~5.}



\clearpage

\begin{deluxetable}{crrrrr}
\tablenum{1}
\tablewidth{0pt}
\tablecaption{Individual Sources}
\tablehead{
\colhead{Feature} & 
\colhead{$\alpha$(B1950)} & 
\colhead{$\delta$(B1950)} & 
\colhead{1720 Line} & 
\colhead{1612 Line} &
\colhead{Velocity} \nl
&\colhead{($00^{\rm{h}} 45^{\rm{m}}00\farcs000$)}&
\colhead{($-25\arcdeg 33\arcmin00\farcs00$)}&
\colhead{(mJy)}&
\colhead{(mJy)}&
\colhead{($\,\textstyle\rm{km~s}^{-1})$}
}
\startdata

1&$05.474\pm0.021$&$42.92\pm0.28$
&$<|\pm 4.2|$&$-10.4\pm1.8$&$248\pm2$\nl 

2&$05.494\pm0.021$&$43.31\pm0.22$
&$9.0\pm1.4$&$<|\pm6.0|$&$262\pm2$\nl

3&$05.546\pm0.018$&$42.15\pm0.18$
&$10.9\pm1.4$&$13.8\pm1.7$&$260\pm2$\nl

4&$05.730\pm0.017$&$40.41\pm0.20$
&$10.6\pm1.3$&$<|\pm6.0|$&$179\pm2$\nl

5&$05.762\pm0.021$&$40.20\pm0.25$
&$8.3\pm1.4$&$<|\pm6.0|$&$166\pm2$\nl

6&$05.763\pm0.025$&$39.67\pm0.27$
&$<|\pm4.2|$&$9.8\pm1.8$&$250\pm2$\nl

7&$05.781\pm0.021$&$39.30\pm0.23$
&$8.6\pm1.4$&$<|\pm6.0|$&$211\pm2$\nl

8&$05.807\pm0.033$&$38.77\pm0.28$
&$8.0\pm1.4$&$<|\pm6.0|$&$217\pm2$\nl

9&$05.841\pm0.027$&$38.82\pm0.23$
&$7.1\pm1.3$&$<|\pm6.0|$&$179\pm2$\nl

10&$05.887\pm0.022$&$38.32\pm0.33$
&$7.0\pm1.4$&$<|\pm6.0|$&$166\pm2$\nl

\enddata
\end{deluxetable}

\setcounter{figure}{0}
\begin{figure}
  \includegraphics{frayer.fig01a.ps} \vspace*{8.in}
\caption{ }
\end{figure}
\setcounter{figure}{0}
\begin{figure}
  \includegraphics{frayer.fig01b.ps} \vspace*{8.in}
\caption{(continued)}
\end{figure}
\begin{figure}
  \includegraphics{frayer.fig02a.ps} \vspace*{8.in}
\caption{ }
\end{figure}
\setcounter{figure}{1}
\begin{figure}
  \includegraphics{frayer.fig02b.ps} \vspace*{8.in}
\caption{(continued)}
\end{figure}

\begin{figure}
  \includegraphics{frayer.fig03.ps} \vspace*{8.in}
\caption{ }
\end{figure}

\begin{figure}
  \includegraphics{frayer.fig04.ps} \vspace*{8.in}
\caption{ }
\end{figure}

\begin{figure}
  \includegraphics{frayer.fig05.ps} \vspace*{8.in}
\caption{ }
\end{figure}

\begin{figure}
  \includegraphics{frayer.fig06.ps} \vspace*{8.in}
\caption{ }
\end{figure}

\begin{figure}
  \includegraphics{frayer.fig07.ps} \vspace*{8.in}
\caption{ }
\end{figure}

\begin{figure}
  \includegraphics{frayer.fig08.ps} \vspace*{8.in}
\caption{ }
\end{figure}

\begin{figure}
  \includegraphics{frayer.fig09.ps} \vspace*{8.in}
\caption{ }
\end{figure}

\begin{figure}
  \includegraphics{frayer.fig10.ps} \vspace*{8.in}
\caption{ }
\end{figure}

\begin{figure}
  \includegraphics{frayer.fig11.ps} \vspace*{8.in}
\caption{ }
\end{figure}

\begin{figure}
  \includegraphics{frayer.fig12.ps} \vspace*{8.in}
\caption{ }
\end{figure}

\begin{figure}
  \includegraphics{frayer.fig13.ps} \vspace*{8.in}
\caption{ }
\end{figure}

\begin{figure}
  \includegraphics{frayer.fig14.ps} \vspace*{8.in}
\caption{ }
\end{figure}

\end{document}